\documentclass[runningheads]{llncs}
\usepackage[T1]{fontenc}
\usepackage{graphicx}
\usepackage{todonotes}
\usepackage{booktabs}
\usepackage{multirow}
\usepackage{graphicx}
\usepackage{subcaption}
\usepackage{comment}

\begin{document}
\title{Designing for Complementarity: A Conceptual Framework to Go Beyond the Current Paradigm of Using XAI in Healthcare}

%

\author{Elisa Rubegni\inst{1,6} \and
Omran Ayoub\inst{2}\and Stefania Maria Rita Rizzo \inst{5,9}  \and
Marco Barbero\inst{3} \and Guenda Bernegger\inst{7} \and Francesca Faraci\inst{4} \and Francesca Mangili\inst{6} \and Emiliano Soldini\inst{7} \and Pierpaolo Trimboli \inst{5, 8}  \and Alessandro Facchini\inst{6}}
\authorrunning{E. Rubegni et al.}
\institute{School of Computing and Communication, Lancaster University, Lancaster, UK 
\and
ISIN, DTI, SUPSI, Lugano, CH
 \and
Rehabilitation Research Laboratory, DEASS, SUPSI, Lugano, CH.
\and
MeDiTech, DTI, SUPSI, Lugano, CH 
 \and
Faculty of Biomedical Sciences, USI, Lugano, CH
\and
IDSIA USI-SUPSI, DTI, SUPSI, Lugano, CH
\and 
CPPS, DEASS, SUPSI, Lugano, CH
\and Servizio di Endocrinologia e Diabetologia, Ospedale Regionale di Lugano, EOC, Lugano, CH
\and Clinic of Radiology, Imaging Institute of Southern Switzerland, EOC, Lugano, CH}

%
\maketitle             
\begin{abstract}
The widespread use of Artificial Intelligence-based tools in the healthcare sector raises many ethical and legal problems, one of the main reasons being their black-box nature and therefore the seemingly opacity and inscrutability of their characteristics and decision-making process. Literature extensively discusses how this can lead to phenomena of over-reliance and under-reliance, ultimately limiting the adoption of AI. We addressed these issues by building a theoretical framework based on three concepts: Feature Importance, Counterexample Explanations, and Similar-Case Explanations. Grounded in the literature, the model was deployed within a case study in which, using a participatory design approach, we designed and developed a high-fidelity prototype. Through the co-design and development of the prototype and the underlying model, we advanced the knowledge on how to design AI-based systems for enabling complementarity in the decision-making process in the healthcare domain. Our work aims at contributing to the current discourse on designing AI systems to support clinicians' decision-making processes.

\end{abstract}


%
\section{Introduction}
Thanks to their efficient capability to analyse vast amounts of complex and diverse data, artificial intelligence (AI for short) systems fuelled by contemporary machine learning techniques are at the forefront of the digital transformation of health systems around the globe \cite{topol2019high}. Notable cases include AI for physician-level diagnostics in dermatology \cite{esteva2017dermatologist} or radiology \cite{laang2023artificial}, for finding optimal treatment strategies for sepsis \cite{komorowski2018artificial} or for identifying patients at risk of cardiac failure in intensive care settings \cite{hyland2020early}. 
However, despite their huge potential, the adoption of AI-based tools raises several ethical and societal issues, one of the main reasons being the seeming inscrutability of their design characteristics and decision-making process \cite{Calster2019,shortliffe2018clinical}. The "black box problem" in AI refers to the challenges and problems that arise because of the opacity of AI systems. For instance, researchers notice that "end users are less likely to trust and cede control to machines whose workings they do not understand" \cite[p.266]{zednik2021solving}, and that "the opacity of AI systems can reduce end users' trust and reliance on using AI-based systems while making critical decisions" \cite[p.1]{haque2023explainable}. 

In line with this is the observation that in reality, when deployed, AI systems are often under-used or not used at all \cite{shah2018big}, one of the reasons commonly put forward being that it is difficult for humans to estimate to what extent to trust recommendations coming from algorithms when no information about their inner behaviour, accuracy\footnote{It has indeed been observed that "people’s trust in a model is affected by both its stated accuracy and its observed accuracy, and that the effect of stated accuracy can change depending on the observed accuracy" \cite{yin2019understanding}.}, or error is given \cite{lee2004trust,hoffman2017taxonomy,jacovi2021formalizing}.

A natural question is, therefore, to understand the exact role opacity is playing in this respect, and which strategies countering it, if any, will eventually enhance trust and enable adoption of AI systems in healthcare.

The view we advocate in this work is that to succeed we need to go beyond the standard XAI methods by aligning with a human-centred perspective on AI (e.g. \cite{Shneiderman2022}). More specifically, in this paper we start to illustrate an approach to co-design `parsimonious evaluative strategies' with users (clinicians), that do not fall into the traps of standard explanatory practices, and that, we hope, encourage trust and the virtuous appropriation of AI, manifested in the generation of complementary performances. Focusing on diagnostic decision support, and embracing a view recently made explicit by Miller in \cite{Miller2023}, the AI decision support, we claim, should merely support the specific diagnostic reasoning of the clinician at stake in the setting under consideration, and thus, in principle, neither be focused on recommending decisions nor on providing explanations for them. 
To do so, we put in place a participatory design approach in which we engaged experts in AI, HCI, and clinicians to co-design and develop an AI prototype for supporting doctors in thyroid disease diagnostics. 
As already mentioned, the design is based on supporting the clinical reasoning but not to directly provide a recommendation. In this paper, we will report about the theoretical background implemented on a case study in which we developed a prototype based on the conceptual framework. 

Our study included \textbf{three iterative phases}: 1. Understanding the Design Space, 2. integrating theory into the design, and 3. co-designing the prototype. 
On one side, we engaged users/stakeholders through a participatory method, while on the other, we developed a high-fidelity prototype using a real dataset. The prototype was partially co-designed with clinicians, who participated as informants to provide feedback.
Our purpose is twofold: 
\begin{enumerate}
\item  to provide a conceptual background for challenging the opacity and encouraging complementarity when AI is used to support decision-making process in health care
\item to advance the knowledge on how to design AI-based systems for the healthcare domain. 
\end{enumerate}
The outcomes of this work aim to contribute to the current discourse on designing AI systems to support clinicians' decision-making processes. We provide concrete examples of potential solutions and advocate for the implementation of a participatory design approach to empower clinicians to actively engage in the process.
The remainder of the paper is structured as follows. In Section \ref{background} we present background on the context of our work while in Section \ref{XAI} we provide the reader with necessary background on explainable artificial intelligence. In Sec. \ref{design} we present our methodological approach for the design process and in Section \ref{casestudy} we describe the case study. In Section \ref{designspace} we elaborate on user needs and Section \ref{prototype} describes the prototype. While Section \ref{codesign} describes the co-design process and Section \ref{conclusion} concludes the paper.

\section{Background}\label{background}
Many among not only the community of computer scientists but also of clinicians and practitioners adhere to the so called \emph{explainability thesis} (ET) \cite{langer2021we,cabitza2023painting}. That is, the hypothesis that explainability is a suitable means for facilitating trust of an opaque AI system in a stakeholder and, thus, making it more acceptable as a decision support tool.\footnote{What actually the cited articles refer to with the thesis is the capability of explaining an AI system's inner behaviour and output. Such thesis is somehow a strong one, and it would be better to formulate it by taking into account the ontology of an AI system, e.g. via the so called levels of abstraction \cite{primiero201610}, and the various forms that opacity may take \cite{facchini2022towards}.}
As such, they are increasingly calling for transparency and explainability to solve the black-box (opacity) problem and build trust in AI systems.  This position is echoed in the EU ethics guidelines for trustworthy AI \cite[p.3]{hleg2019ethics}, that considers transparency and explainability as "crucial for building and maintaining users' trust in AI systems". 

The ET has, however, been challenged. In addition to pointing out that it is not clear what the explainability or interpretability of AI system actually amounts to,  one counter-argument starts by noticing that "[as] counter-intuitive and unappealing as it may be, the opacity, independence from an explicit domain model, and lack of causal insight associated with some of the most powerful machine learning approaches are not radically different from routine aspects of medical decision-making. Our causal knowledge is often fragmentary, and uncertainty is the rule rather than the exception. In such cases, careful empirical validation of an intervention's practical merits is the most important task. When the demand for explanations of how interventions work is elevated above careful, empirical validation, patients suffer, resources are wasted, and progress is delayed" \cite[p.18]{London2019}. 
Thence, according to London \cite{London2019}, what is needed for enabling trust and adoption is not so much explainability, intended as the capability of providing explanations for the behaviour or output of an AI system, but rather accuracy and reliability. 
Indeed, current explainability techniques, mainly stemming from the eXplainable (XAI) research programme \cite{Ali2023}, not only fall short in their original aim but they might have unintended negative effects. For instance, aligned with London's view, despite some positive results (see e.g. \cite{tonekaboni2019clinicians,diprose2020physician,jabbour2023measuring}), experiments have shown that sometimes accuracy is more important for user trust than explainability, and that adding an explanation for a recommendation can potentially harm trust when the fidelity of the explanation is low \cite{papenmeier2019model,papenmeier2022s}. In addition, explanations of recommendations can lead to automation bias and over-reliance on AI systems (the mere presence of an explanation often already increases trust), to accept incorrect decisions and explanations without verifying whether they were correct, and thus to unjustified (unwarranted) trust in AI recommendations, or to cause reasoning errors such as confirmation bias and thus to groundlessly increase confidence in one’s own decision.\footnote{See \cite{bertrand2022cognitive} for a review of existing literature on this aspect. 
}
This is the reason why, as explained in \cite{cabitza2023painting}:

\begin{quote}
"providing AI with explainability [\dots] is more akin to painting the black box of inscrutable algorithms [\dots] white, rather than making them transparent. What we mean with this metaphoric statement is that XAI explanations do not necessarily explain (as by definition or ontological status) but rather describe the main output of systems aimed at supporting (or making) decisions".   
\end{quote}
As a consequence, using explanations does not necessarily enable achieving complementarity. That is, it doesn't necessarily make a "hybrid team" composed by a human and an AI to take better decisions than humans or AIs alone. Actually, the only situation where the human-machine collaboration via explainability outperforms people alone is when the accuracy of the underlying AI model is higher than human accuracy \cite{bansal2021does}, but still the performance of the hybrid team is lower than the one by AI alone \cite{buccinca2021trust}. Hence, one can argue with \cite{ghassemi2021false} that, in the absence of suitable explainability methods, it is better and safer to rather advocate for rigorous internal and external validation of AI systems (a task for which current XAI methods may actually be very useful). 
What we should look for is not a full `transparency' of the trained algorithm, but rather a form of design transparency, that is "the adequate communication of the essential information necessary to provide a satisfactory design explanation of such a system" \cite{Loi2021}, such as the data's origin and the type of data used in training\footnote{See e.g. the case described in \cite{kamulegeya2019using}.} (including how risks of biases have been tackled), the goal of the algorithm\footnote{In particular the intended functionality assigned to the trained model by the designers.}, and thus its adequacy (applicability) to the context is supposed to be deployed and used, and as well as characteristics such as its validation and accuracy (see \cite{arbelaez2022re} for a discussion on these points). 
Unfortunately, it has been shown, e.g. by \cite{bansal2021most}, that being the most accurate model does not necessarily imply it to be the 'best team-mate in the room'. The reason simply being that a collaborative context "puts additional demands on participants that extend beyond individual performance on tasks, such as ability to complement and coordinate with one’s partner". In fact, accuracy on training data is \emph{not} equal to accuracy on unseen data when the system is deployed and that the accuracy of a decision support system is \emph{not} the accuracy of decision making (see e.g. \cite{cabitza2021err}). Moreover, remember that we are considering a situation in which, due to the opacity of the system, the only information that humans are relying on to judge the correctness of an AI recommendation is the decision maker's own expertise and background knowledge.\footnote{We can also assume that information contributing to design transparency are included in the background knowledge.}
Given this, even though the AI is better than the human (keeping in mind the previous caveat on such a claim), it is well known that people still tend to under-rely and thus ignore the recommendation of the machine because of \emph{unwarranted distrust} or to over-rely and thus incur in the acceptance of wrong decisions due to \emph{unwarranted trust} \cite{miller2019explanation,gunning2019darpa,jacovi2021formalizing,zhang2020effect,bansal2021does,gajos2022people,sivaraman2023ignore}. We thus have reached a paradoxical conclusion: opacity seems to hinder the virtuous adoption of AI in the healthcare sector, but tools from XAI may have a counterproductive effect. At the same time, simply providing AI recommendations (perhaps with some additional information, e.g. accuracy) neither seems to promote adoption and complementarity. The approach advocated in this paper aims at countering this paradoxical situation by asking for a change of paradigm in the design and use of XAI in diagnostics.

\section{XAI: A Three Pillars Conceptual Framework}\label{XAI}
XAI is a rapidly evolving area within the field of AI that seeks to clarify the decision-making processes of AI systems and opaque ML models, especially in critical applications such as healthcare and finance. Specifically, XAI aims to bridge the gap between the inherent complexity of advanced AI algorithms and the need for transparency in understanding how these systems arrive at specific outcomes. By providing insights into the internal workings of AI models, through a broad set of families of explanations, XAI aims to enhance the interpretability of results and to foster trust among end-users, regulators, and stakeholders, ultimately promoting responsible and ethical AI deployment.

The explanations generated using XAI techniques can be either local, explaining model's decision for an instance (i.e., how the model arrived to a decision for a particular instance), or global, explaining the overall working of the underlying machine learning model. In our prototype, and since we are simulating a use case for diagnostic settings, we leverage local explanations. Specifically, our prototype provides, for the instance of interest (i.e., for the instance being analyzed by the clinician), a set of local explanations that assist the clinician in the diagnostic. In particular, we leverage three families of explanations, namely, \emph{Feature Importance Explanations}, \emph{Counterexample Explanations} and \emph{Similar-Case Explanations}. In the following, we provide a short description of these families of explanations.  

\paragraph{Feature Importance}
(or, feature attribution) refers to the measure of the impact or contribution of individual input features in a machine learning model's decision-making process \cite{ribeiro2016should,lundberg2017unified}. It helps to identify, based on the XAI technique employed to compute the importance of the features, which features have the most significant influence on the model's predictions. The feature importance of a specific feature for a given data sample being analyzed (i.e., the decision of the ML model for the data sample being explained) can be as either positive or negative. A positive importance value signifies that the feature supports the decision favoring the considered class, while a negative importance value suggests that the feature does not contribute to the decision in favor of the class under consideration.

When predicting thyroid disease, conducting a feature importance analysis can provide valuable insights into the factors that significantly influence the model's diagnosis. This analysis helps identify the key contributors, such as age, gender, or thyroid hormone levels, that play a crucial role in either supporting or contradicting a particular diagnosis. For instance, if the feature importance analysis indicates that age is the most influential factor, it suggests that the model heavily relies on age information when making its prediction. This insight into feature importance aids in understanding the model's underlying logic, guiding users to identify how the ML model reached its decision.

\paragraph{Counterexample Explanations} 
are a type of explanation that explores what could have happened in a given situation if certain factors or events had been different \cite{holzinger2019causability,ribeiro2016should,verma2020counterfactual}. In other words, they involve constructing hypothetical scenarios or \emph{counterfactuals} to allow users to understand how changes in specific variables (input features) in a specific data sample might have led to different outcomes \cite{wachter2017counterfactual,li2022interpretable}. 
In this context, counterfactual explanations are often used to explain the predictions or decisions made by a ML model. By identifying the key features or inputs that influenced the model's output, one can create counterfactual instances where those features are modified to observe how the model's prediction would change. This helps users (and developers) better comprehend the decision-making process of ML models and gain insights into their behaviour and, in fact, counterfactual reasoning have already proven to improve the interpretability of ML models in several domains such as healthcare \cite{prosperi2020causal}.

Within the context of our specific use case, the extraction of counterfactuals provides the user, notably the clinician, with a valuable array of options. These options delineate how alterations in input features, encompassing the patient's information and, for example, test measurements, could potentially impact the ultimate the diagnosis.

\paragraph{Similar-Case Example Explanations} involve presenting instances from the dataset that closely resemble the input data (i.e., the data record currently under investigation) such as to to illustrate how analogous cases influenced the model's decision. Such explanations can help identifying instances in the dataset that share similarities with the input data, and that have been given the same label \cite{chen2023understanding}. This approach is particularly valuable in providing concrete and relatable examples that can enhance the interpretability of machine learning models. 

In the cases under consideration in this work, these three explanatory approaches will constitute the basis of our framework.

\section{Designing with the Users: Methods, Data Collection and Analysis}\label{design}
Our design research aligned with a human-centred perspective on AI (see e.g. \cite{Shneiderman2022}) and grounded the idea that the AI decision support should be designed as in such a way that the human decision maker maintains control over which hypotheses to explore. Thus, we implemented a design approach that aims at developing a system that aligns with the idea that is the decision support system should be designed to explicitly support the abductive reasoning process, that is in the medical domain the (differential) diagnostic type of reasoning of a clinician. In order to do this, the project followed a Participatory Design approach (PD) \cite{simonsen2012routledge} by involving the stakeholders (e.g. hospital clinicians, researchers and technicians) in every step of the design process of the system.  In this work, we have involved users and stakeholders in different ways and their participation had different degrees of engagement within the process according to the people's roles and the phase of the project. Specifically, across the research project they have been involved as: users, testers, informants, and co-design partners \cite{druin2002role}. As users or testers, the target group is observed during their normal activity using the tools we provided. They are inquired about their everyday workflows, tools, and tasks. Later on, they are asked for acceptability and/or usability of an early version of the tool. As informants, users are solicited for input and feedback. As, co-designers the users are considered as equal partners in the design process. Users are actively engaged and invited to provide input starting at an early stage of design, preceding the development of a fully working prototype \cite{druin2002role}. Drawing upon the PD potential to foster user adoption, our study is grounded in this approach \cite{frauenberger2015pursuit}. Indeed, the PD approach increases the likelihood of stakeholders and users adopting the tool, as well as the team better understanding the context, wishes, and needs thus increasing the chances of really supporting their activities. PD has been successfully and widely used within this domain (e.g. \cite{desmet2016participatory} \cite{noorbergen2021using}). 
Our approach consists of three iterative phases: \textbf{1. Understanding the Design Space}, \textbf{2. integrating theory into the design}, and \textbf{3. co-designing the prototype}. 
Overall, we engaged 8 people with different domains of expertise: two clinicians, three medical technicians, a user experience researcher, an expert in AI in healthcare, and an expert in ML in healthcare. Through their active engagement, participants contributed in defining the problem space, eliciting the system requirements, and exploring the different designs presented as low-fi and hi-fi prototypes.  
In the project, we developed a hi-fi prototype and built a ML model by using a real data set. The creation of a hi-fi prototype allowed us to challenge our scientific hypothesis and the conceptual framework by directly engaging the clinicians as informants into a co-design session. We collected data by transcribing notes and taking pictures of the context in phase one. We analysed data by defining the design space and the requirements that directly feed to design of the prototype. In the co-design stage, we recorded the session, transcribed notes, and later analysed the video.  

\section{The Case Study}\label{casestudy}
In this paper, we report on a specific case study developed in collaboration with the \textit{Ente Ospedaliaro Cantonale} (EOC - the institution that manages, coordinates, and integrates a network of public hospitals in Ticino, Switzerland). 

We present the activities that we conducted with the users and stakeholders to \textbf{identify the problem space}, subsequently we illustrate the \textbf{integration of the theory into the design of the prototype}, and then we present the first round of \textbf{co-design} in which a clinician was involved as informant to provide feedback on the prototype \cite{druin2002role}. 

To recruit participants, we used a convenience sampling approach. Leveraging the researchers’ network, we contacted a few people who work in the local research hospital. For the first phase (identifying the design space) we recruited on a voluntary basis a team of radiologists (technicians, clinicians, and researchers). This team was willing to collaborate with us and aligned with our vision on AI in the decision-process as a complementary tool to help in making decisions. 
Overall, we involved three technicians, that have been in the role for over 10 years: a clinician who leads the local research unit in the Clinic of Radiology at the EOC, and is also professor at the Faculty of Biomedical sciences of the University of Southern Switzerland, where she is involved in many research and teaching activities related to imaging and AI, and a researcher which is part of this team.  
For the third phase (co-design of the first hi-fi prototypes), we recruited another clinician in the domain of endocrinology (and also a scholar of the University of Southern Switzerland) who participated on a voluntary basis, reaching out to the team that had joined the first phase (at the Clinic of Radiology). This user is the head of the Endocrinology Department at the EOC and acted as co-designer and provided input on a prototype based on a Thyroid Disease dataset. The rationale behind recruiting individuals with diverse specialties is to challenge our concepts to scrutiny across various health domains.  

\subsection{Threats to Validity}
In our case study, we involved a limited number of users and stakeholders recruited using convenience sampling through the university network. However, this is not unusual in participatory design research. Indeed, this approach allowed us to delve deeply into each participant's practices, needs, and wishes. As a result, we were able to provide a complete and exhaustive overview of the clinicians and technicians involved. Therefore, engaging a larger sample would not have allowed at this stage of the project to get such a deep understanding of the users. Moreover, this approach allowed us to create a strong connection with the participants that lead to a long term partnership. However, this should be considered when generalising the results. To strengthen our approach and address potential external validity issues in the co-design phase, we have recruited a clinician with a different specialty than the other participants involved in phase 1. This helped strengthen our conceptual model by subjecting it to challenges across various health domains (Diagnostic Imaging and Endocrinology), diseases (Pulmonary nodules and Thyroid Disease), and types of data (images and text). Following we report a summary of the outcomes of the first phase. 
\section{Phase 1: Understanding the Design Space}\label{designspace}
In the first phase of the project we investigated the design space. We involved three technicians, one clinician, and one researcher who were inquired about their practices, the tools used, their workflow, including the people who are part of the decision making process while formulating a diagnosis. In order to explore these aspects, three authors conducted four contextual inquiry interviews on their laboratories and offices (see the Annex). The focus of these interviews was on investigating their practices and the potential issues and opportunities to address in designing the new system. We interviewed them separately to enable each participant to focus deeply on their own activities and tasks.
We interviewed the technicians and the researcher one time. While, to delve deeper on their activities, we inquired the clinicians in three rounds.
Each inquiry lasted about 90 minutes and was based on a semi-structured interview in which the users could talk over their everyday practices in patient examinations and making options about a diagnosis. Being in their offices and laboratories allowed them to show us the tools that they use everyday, including the software. During the second interview round we also explained our hypothesis with the purpose of investigating whether they would be open to use an AI-based system grounded in our conceptual model. 

In \textbf{the first round}, the participants agreed it is about providing counterfactual examples while the clinician is evaluating the different options. The type of counter-examples varies depending on the data and the domain. For instance, if the system is tailored for radiologists, examples may consist of images accompanied by brief descriptions. Conversely, for endocrinologists, the data might be numerical and textual in nature.
From the interviews, it emerged that, at the moment, technicians do not provide any insight on the diagnosis thus they might not need this type of support. While, for clinicians, it would be extremely relevant to have that type of support. 

In \textbf{the second round}, they demonstrated to be extreme positive and enthusiast about having a set of counterexamples instead of a direct recommendation, unlike other AI-based systems they have previously used. They mentioned multiple times during the interviews that using counterfactual reasoning is the strategy they usually applied when they have to make a decision. Usually, they asked for second opinions from the colleagues who provide insights on the basis of their experience. Hence, they highly appreciated the prospect of an AI-based system that could complement their reasoning and support their decision-making process by offering Counterfactual Explanations and Similar-Case Explanations. Such support is particularly desirable to them, especially if it is backed by a robust and extensive dataset. During this second round of interviews, we co-identified a set of examples (pro and against) that supported the counterfactual reasoning and lead to improvements in the interpretability of ML models. We focused mainly on Pulmonary Nodules and they provided us five potential diseases to be considered in a differential diagnosis process (congenital, inflammatory, neoplastic, vascular, and miscellaneous).

In \textbf{the third round}, we discussed with the participants the number of examples to provide (both for and against), and they expressed the preference to include one example for each of the five potential diagnoses. Connected with this point, clinicians helped us to identify the specific features that influence their decision and their importance. These factors aimed at influencing the output provided by the ML model. As well as these changed according to the domain. We identified these examples by combining the features provided by the dataset with the expertise of the clinicians. For instance our stakeholders identifies 20 features that will be used to implement the prototype. Furthermore, we asked participants to bring concrete examples of a set of features, counterfactual, and similar-case explanations that we will used to build the model. The participants created an anonymised data set that we will use in the next step of the project for developing a prototype for supporting Diagnostic Imaging decision-making process. The patient information were anonymised to ensure the protection of privacy and confidentiality of their data.



\section{Phase 2: Integrating the Theory into the Design}\label{prototype}
In the second phase, we integrated the theory into the design by developing a prototype. The prototype is a high-fidelity interactive artifact based on a machine learning model trained on a real dataset. This second phase runs in parallel with the first one, however they feed one other. Specifically, phase one helped consolidating our conceptual framework. In this phase, the three-pillars framework was implemented into a prototype by using a real dataset of thyroid disease diagnosis domain. We opted for a different domain (endocrinology) than phase one (radiology) for two main reasons: to challenge our model, and because we found a rich dataset that could be used to develop and train a machine learning model, which was currently lacking in radiology. In this section, we explain how the model was built and what the prototype and interaction design look like. 

\subsection{ML Model Development and Explainers}
The thyroid disease diagnosis problem can be formulated as a multi-class classification problem. The diagnosis task (i.e., the classification of a given data record) involves distinguishing between hyperthyroid conditions, hypothyroid conditions, and negative cases (i.e., neither hyperthyroid nor hypothyroid disease). 
The dataset used in our study is constructed by merging six datasets from the Garavan Institute of Medical Research.\footnote{https://www.garvan.org.au/} The dataset has been made publicly available and is currently available for download from the UCI machine learning repository.\footnote{https://archive.ics.uci.edu/dataset/102/thyroid+disease}
We have preprocessed the original dataset by eliminating data records (data samples) with missing values. After this step, we are left with a dataset of 7142 data records. The records are distributed among the three classes as follows:
\begin{itemize}
    \item Negative (also referred to as class 0) consists of 6385 data records (89.4\% of the overall records in the dataset)
    \item Hyperthyroid (class 1) consists of 582 data records (8.15\% of the overall records in the dataset)
    \item Hypothyroid (class 2) consists of 175 data records (2.45\% of the overall records in the dataset)
\end{itemize}
The input data consisting of features (variables) with either Boolean values or numerical values. Tab. \ref{features} reports the list of features along with a brief description. 
\begin{table}[htbp]
    \centering
    \caption{List of input features pertaining to patient information}
    \begin{tabular}{lcc}
        \toprule
        \textbf{Variable} & \textbf{Type} & \textbf{Description} \\
        \midrule
        age & Integer & Age of the patient \\
        sex & Boolean & Sex of the patient \\
        on\_thyroxine & Boolean & Whether patient is on thyroxine \\
        on\_antithyroid\_meds & Boolean & Whether patient is on antithyroid meds \\
        sick & Boolean & Whether patient is sick \\
        pregnant & Boolean & Whether patient is pregnant \\
        thyroid\_surgery & Boolean & Whether patient has undergone thyroid surgery \\
        I131\_treatment & Boolean & Whether patient is undergoing I131 treatment \\
        query\_hypothyroid & Boolean & Patient believes they have hypothyroid \\
        query\_hyperthyroid & Boolean & Patient believes they have hyperthyroid \\
        lithium & Boolean & Whether patient takes lithium \\
        goitre & Boolean & Whether patient has goitre \\
        tumor & Boolean & Whether patient has a tumor \\
        hypopituitary & Float & Hyperpituitary gland status \\
        psych & Boolean & Whether patient has psych \\
        TSH & Float & TSH level in blood from lab work \\
        T3 & Float & T3 level in blood from lab work \\
        TT4 & Float & TT4 level in blood from lab work \\
        T4U & Float & T4U level in blood from lab work \\
        FTI & Float & FTI level in blood from lab work \\
        \bottomrule
    \end{tabular}
     \label{features}
\end{table}
To build the ML model for thyroid disease classification, we train and test an extreme gradient boosting (XGB) model in a supervised manner following a 10-fold cross validation with 80-20 train-test split. The XGB model was optimized through hyper-parameter tuning and the resulting model achieves an average accuracy of 0.99 across the 10 folds. The model achieves a precision\footnote{The precision represents the ratio of correctly predicted instances of that class to the total instances predicted as that class}, recall\footnote{The recall represents the ratio of correctly predicted instances of that class to the total instances of that class in the dataset} and F1-score\footnote{The F1-score is the harmonic mean of precision and recall} values that range between 0.8 and 0.99 for Negative and Hyperthyroid class while values that range between 0.75 and 0.97 for the Hypothyroid class. The results indicate that the Hypothyroid class represents a challenge for the ML model, however this is expected considering the relatively low number of data records present for that class. It is important to note that the classification performance does not directly influence the design of our prototype. Instead, we report them as they serve as a foundational benchmark for our ongoing and future studies in this domain.
To extract explanations, we leverage two XAI techniques. To compute feature importance, we use Local Interpretable Model-Agnostic Explanations (LIME) framework \cite{ribeiro2016should}. LIME works by approximating the decision boundary of the black box ML model in the vicinity of a specific instance of interest (the data record being explained), thus generating what is referred to as \emph{locally faithful explanations} \cite{ribeiro2016should}. To achieve this, LIME employs a two-step process: perturbation and approximation. In the perturbation step, LIME samples a set of instances around the data point under consideration and perturbs them by introducing slight modifications. Subsequently, these perturbed instances, along with their corresponding model predictions, are used to train an interpretable surrogate model such as, e.g., linear model. Feature importance is then computed based on the weights assigned to each feature in the surrogate model, reflecting the contribution of individual features to the model's decision within that specific locality. This localized interpretability enables a more nuanced understanding of the model's behavior and enhances transparency, as it allows the user to understand which features (or factors) are most influential for a specific model's decision. The effectiveness of LIME has been demonstrated across various domains, making it a valuable tool for feature importance computation and model interpretation \cite{hailemariam2020empirical,jeyakumar2020can,pawar2020incorporating,manresa2021advances}. Yet, we note that our prototype can rely on any XAI technique for computing feature importance. 

To generate similar-case example explanations and counterexamples, we use DiCE (Diverse Counterfactual Explanations) \cite{mothilal2020explaining}. DiCE is a method that employs optimization and heuristic approaches aimed at producing a set of counterfactual explanations that are close, in terms of their proximity measure, to the original input data record (i.e., the data sample being explained). More specifically, for a given input data point, which is, in our case, the data record whose decision is to be explained, we employ DiCE to generate a set of counterfactuals or, in other words, counterexamples that are data samples classified as a different label (class) by the ML black box and that exhibit minimal differences in both the number of features and the extend to which features differ compared to the original data record. 
To generate similar-case example explanations, we constrain DiCE to generate data samples that differ minimally with respect to original data record however belong to same class (i.e., are not labeled different by the ML model). It is important to highlight that any counterfactual explainer can be used in our prototype, and it is not constrained solely to DiCE. Moreover, we note that, while DiCE has proven its efficacy in generating interpretable and actionable counterfactuals \cite{guidotti2022counterfactual,jia2022role}, we plan to consider and eventually compare several counterfactual explainers in future work. 

\subsection{Prototype Interface and User Interaction}
\begin{figure}
  \centering
  \begin{subfigure}[b]{0.41\textwidth}
  \centering
    \includegraphics[height=4.6cm, width=0.75\textwidth]{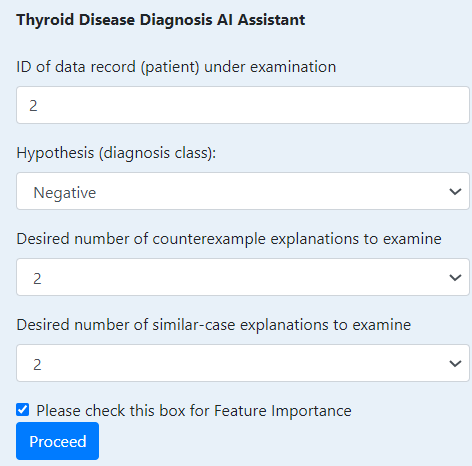}
    \caption{\footnotesize{Screenshot of the page the users encounter when using the prototype showing the list of input parameters that the user has to provide.}}
    \label{fig:interface1}
  \end{subfigure}
  \hfill
  \begin{subfigure}[b]{0.5\textwidth}
  \centering
    \includegraphics[height=6cm, width=0.7\textwidth]{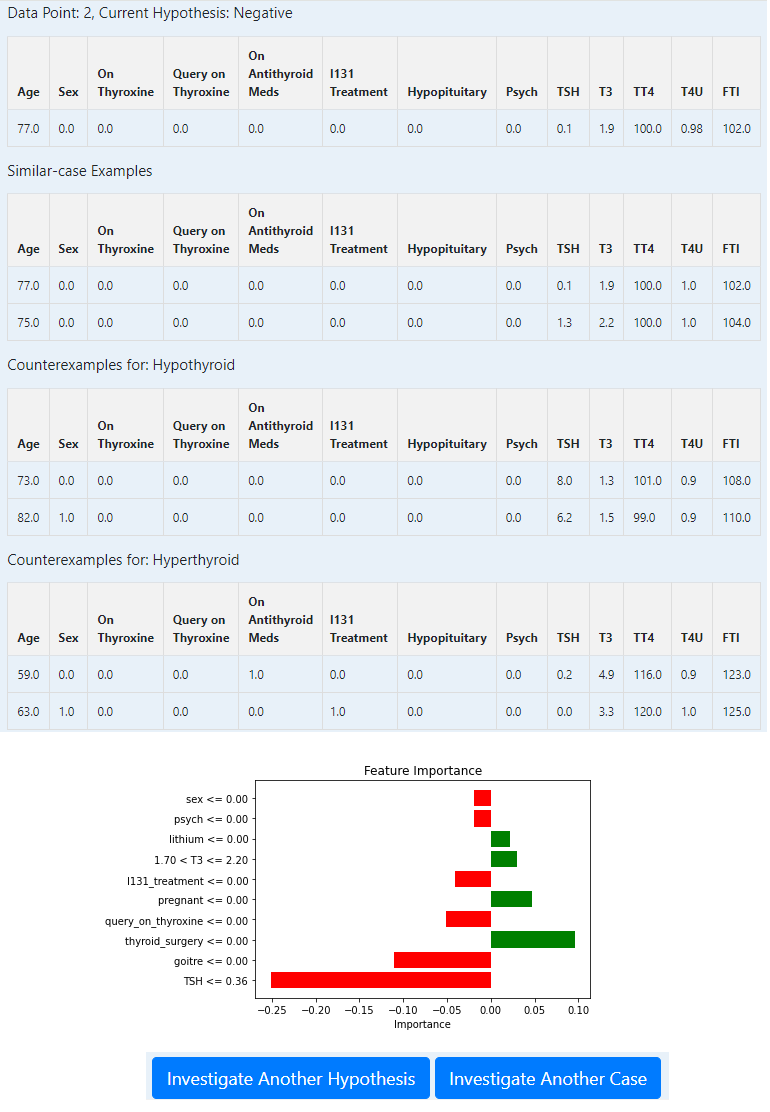}
    \caption{\footnotesize{Screenshot of the page that shows the system's outcome.}}
    \label{fig:interface2}
  \end{subfigure}
  \caption{Two screenshots of the prototype interface.}
  \label{fig:combined}
\end{figure}

We now present a description of user's interaction with the prototype through the developed interface. 
The process consists of four steps. We assume that prior to interacting with the tool, the clinician examines the case at hand and formulates a hypothesis of the diagnosis of the case as belonging to one of the three classes, i.e., either \emph{hyperthyroid}, \emph{hypothyroid}, or \emph{negative}. Once the clinician has formulated a hypothesis, the clinician engages with the tool by inputting a set of parameters.
Figure \ref{fig:interface1} shows the initial page that users encounter. The input parameters required by the user are the following:
\begin{itemize}
\item ID of the data record (or, patient) under examination in the field \emph{ID of data record (patient) under examination}. 
\item The hypothesis of the diagnosis, specifying the class that the clinician believes the case under examination should be classified as in the field \emph{Select your hypothesis (class)}. The user has to select one of the three possible diagnosis classes. 
\item The desired number of counterfactual explanations the clinician wishes to investigate for each of the other two classes of diagnosis. Our system allows users to choose between 0 (no such explanations) and 10 counterexamples for each of the other diagnosis classes.
\item The number of similar-case example explanations the clinicians wishes to investigate for each of the other two classes of diagnosis. Our system allows users to choose between 0 (no such explanations) and 10 similar-case example explanations.
\end{itemize}
In addition to these inputs, the user can select whether or not to have the tool compute and display feature importance explanation by checking the relative check-box present on the initial page. Finally, the clinician proceeds by clicking on \emph{proceed}. After that, the tool provides the user with the outcome. Fig. \ref{fig:interface2} shows as example of the outcome of the tool for a given data record, which consists of the following: 
\begin{itemize}
    \item The ID of the data record under examination and the hypothesis selected by the user.
    \item The data record under evaluation, including the features and their values. 
    \item The set of similar-case example explanations. The example shows 3 explanations.
    \item The set of counterexamples for each of the other two classes. The example shows 3 explanations for each of the two classes. 
    \item The feature importance figure showing which features exhibited positive or negative influence towards the hypothesis selected by the user.
\end{itemize}

In the example shown in the figure, the system's outcome shows that similar-case explanations can be obtained by slightly altering the values of \textit{T3}, \textit{T4U} or \textit{FTI} and also shows that the \textit{TSH} value can be around 1.3 as opposed to 0.1 in the second similar-case explanation. In terms of counterexample explanations, the outcomes reveals that data records labeled as \emph{Hypothyroid} can be obtained by drastically changing the value of \textit{TSH} (from 0.1 to 8.0 for the first counterexample explanation, or to 6.2 for the second counterexample explanation) while minimally altering values of other features (\textit{Age} was slightly reduced (from 77 to 73) and the features \textit{T3}, \textit{T4}, \textit{T4U} and \textit{FTI} undergone relatively small alteration). Similarly, the counterexample explanations for the \textit{Hyperthyroid} also reveal how feature values can be altered to obtain the relative class. 

After analyzing the system's outcome, the clinician can take a decision with full autonomy, if decided on the diagnosis, or proceed by testing another hypothesis for the case under investigation by clicking on \textit{Investigate Another Hypothesis} button. 

\section{Phase 3: Co-designing the Prototype}\label{codesign}
In this third phase, we have conducted two workshop sessions in which we engaged an interdisciplinary group of experts in four key domains: user experience, AI in health care, ML in health care, and endocrinology. 

During \textbf{the first workshop}, the conceptual model was explained to the clinicians who had the chance to provide feedback and ask questions. In that session we explored also the clinical domain(endocrinology) and its specific challenges and opportunities. We delved into his everyday practices, and on how he deals with the differential diagnosis process. It was shown also the data set on thyroid diseases and we discussed with him the features and, finally, selected the most relevant (Table \ref{features}).

The \textbf{second workshop} was focused on presenting and interacting with the prototype. This hands-on experience allowed him to articulate his needs and expectations for the prototype as well as to envision new functions. In this sense, the participants acted as informants \cite{druin2002role}. This session was not intended for evaluating the technical solution; instead, its purpose was to demonstrate the application of the theoretical framework into a high fidelity prototype. The session was conducted online via Teams and recorded to better analyse the results. Two co-authors facilitated the session. First, the participants were asked to perform a series of tasks and provide feedback by thinking aloud. This technique allowed participants to verbally express thoughts about the interaction experience, including their motives, rationale, and perceptions. Subsequently, they engaged in a brainstorming session in which they were asked to contribute to producing ideas to better implement the concept (and the theory) into the prototype to better support their  practise. The session lasted 90 minutes. During the workshop the participants discussed the use of the prototype into the clinical practices at large and in details by looking at each items implemented. Overall the clinicians really liked the concepts on which the prototype was based. He expressed a positive opinion about its usage on his own practices as well as in suggesting it to other colleagues. During the brainstorming the team explored how the provided functionalities could be extended and whether the information provided could be improved.  For what concern the features during the brainstorming we refined their types and order to better assist clinicians in comparing the case with the counterexamples. For instance, the TSH value is given priority after the age and gender of the patient. In addition, there is no point of asking if the features need to be shown or not as these are considered relevant in all scenarios, and the system should provide them regardless. Participants agreed that the system should offer a choice between 'more important' or 'less important'.
Regarding the number of counterfactual explanations and similar-case explanations provided by the system, the clinician explained that this depends on the type of diagnosis he is evaluating. For instance, in the case of euthyroid the clinicians does not need to have many and just 3 of them would be enough. In other diagnostic cases, such as hyperthyroidism and hypothyroidism, where a disease might be present, the system should suggest a higher numerical value. This would provide better support for abductive reasoning to clinicians, enabling them to explore various options and compare data from other patients.  Thus, the participants agreed that a good solution was to provide a default number which is smaller (e.g. 3) in the case of euthyroid and larger (e.g. 5) in the case of hyperthyroidism and hypothyroidism. Then, eventually, the clinicians would have the option to modify the default number as needed. As output of the session we also produced a set of screenshots of the prototype with recommendations and additional features co-created by the participants. These will guide the next step of the design process, leading to a new version of the prototype.

\section{Conclusion and Future Work}\label{conclusion}
Inspired by \cite{Miller2023}, in this work we presented a conceptual framework to go beyond the current paradigm of using XAI in healthcare, with a specific focus on clinical diagnostic. The advocated paradigm change is based on the view that the design of a AI decision support systems has to focus on supporting the specific diagnostic reasoning of the clinician at stake in the setting under consideration. 
More specifically, based on three explanatory techniques -- feature importance, counterexample explanations (based on counterfactuals), and similar-case explanations --, we illustrate a participatory design approach in which we engaged experts in AI, HCI, and clinicians to co-design and develop an AI prototype supporting the clinical reasoning without directly providing neither a recommendation nor, a fortiori, an explanation for it.  
The results of this iterative process showed how the conceptual framework could be first implemented in a ML model and then concretized into a hi-fi prototype, and how the users/stakeholders needs could be integrated and harmonised into the designed solution. During the co-design in the brainstorming session, the initial design ideas have been refined and this helped to move on to the next step. The limited sample we engaged in the study enabled us to delve into their knowledge, practices, resources and tools that they use to form the cognitive process leading to a clinical diagnosis. This immersive process allowed us to design a hi-fi prototype that embodied both the theoretical framework and the users needs. Leveraging on these result a new version of the prototype will be developed by using the anonymised data set produced by the research units at the EOC that we engaged in the first phase of the project. This prototype will be evaluated with a larger sample of clinicians, experts in the concerned domain. The purpose of this follow up study will be to assess the ability of the prototype to support the specific diagnostic reasoning and, thus, to verify that complementarity is achieved. 
To conclude, with this work we hope that, in addition to advance the knowledge on how to design AI-based systems for the healthcare domain, we also contribute to the current discourse on promoting virtuous adoption, trust and best practices related to such systems. 

\section*{Acknowledgements}
The work presented in the paper is part of the exploratory research programme \emph{Best4EthicalAI}, financed by SUPSI. We would like to thank SUPSI for supporting our research, as well as all the individuals who have been involved in the study. Participants gave their consent to participate voluntarily in the case study that was approved by the SUPSI ethical committee. Without them, this work wouldn't have been possible. A special thanks to Gail Collyer-Hoar for her invaluable job she did in proofreading of the paper. 


%
%
 \bibliographystyle{splncs04}
 \bibliography{Designingforcomplementarity}

 \end{document}